\documentstyle[11pt,axodraw,a4]{article}
\newcommand\Li{\mathop{\rm Li_2}\nolimits}
\newcommand\order{\mathop{\cal O}\nolimits}

\newcommand\setscale[1]{\def\axoscale{#1 }}

\newcommand{\gsim}{\stackrel{\lower.7ex\hbox{$>$}}
    {\lower.7ex\hbox{$\sim$}}}
\begin{document}

\thispagestyle{empty}

\setcounter{page}{0}

\begin{flushright}
PSI-PR-93-22\\
ZH-TH 30/93\\
TTP 93-33\\
April, 1994\\
REVISED
\end{flushright}

\vspace*{\fill}

\begin{center}
{\Large\bf Unstable particles in One Loop Calculations}\\
\vspace{2em}
\large
\begin{tabular}[t]{c}
Andr\'e Aeppli$^{a\dagger}$\\
Geert~Jan van Oldenborgh$^b$\\
Daniel Wyler$^c$\\
\\
{$^a$ \it Institut f\"ur Theoretische Teilchenphysik}\\
{\it Universit\"at Karlsruhe}\\
{\it Kaiserstr.\ 12, D-7500 Karlsruhe 1, Germany}\\
{$^b$ \it Paul Scherrer Institut, CH-5232 Villigen PSI, Switzerland}\\
{$^c$ \it Institut f\"ur Theoretische Physik}
{\it Universit\"at Z\"urich}\\
{\it Winterthurer Strasse 190, CH-8057 Z\"urich, Switzerland}\\
\end{tabular}
\end{center}
\vspace*{\fill}

\begin{abstract}\noindent
We present a gauge invariant way to compute one loop corrections to processes
involving the production and decay of unstable particles.
\end{abstract}
\vspace*{\fill}
{\footnotesize$^\dagger$ Current mail address:\\
Institut f\"ur Theoretische Physik, Universit\"at Z\"urich, Winterthurer
Strasse 190, CH-8057 Z\"urich, Switzerland}

\newpage


\section{Introduction}


In this article we describe a gauge invariant way to compute the one loop
corrections to processes involving the production and decay of unstable
elementary particles, such as the $W$, $Z$ and Higgs bosons or the top quark.
All these particles have widths which are (possibly) a sizable fraction of
their mass ($\Gamma/m \approx 1/40$ for the $W$ and $Z$ bosons).  A
fixed-order amplitude in perturbation theory respects the symmetries of the
theory, like gauge invariance.  However, in a reaction which involves the
production and decay of these unstable particle this amplitude contains a
non-integrable infinity due to the propagator $1/(p^2-m^2)$.  We are thus
forced to take into account higher order diagrams: the (one-particle
reducible) self energy graphs of the unstable particle.  Resumming these gives
rise to a pole which now is off the real axis, and the amplitude is finite.
However, one can no longer assume that this amplitude is still gauge
invariant, as one has mixed different orders in perturbation theory.
Experience has shown that gauge dependent terms can sometimes be much larger
than the physical answer, so it is necessary to investigate this problem.

There are many ways in which the gauge invariance of the amplitude can be
restored.  The easiest way to circumvent the problem is to use the narrow
width approximation, that is, to factorize the process into production and
decay of the unstable particle.  This is often not good enough when the width
of the unstable particle is sizable.  Another approach is to introduce the
width in the Lagrangian, by adding and subtracting suitable terms, thus using
a complex width everywhere.  This is a problematic procedure, as we will
discuss in section \ref{sec:kin}.  Finally, it is often possible to obtain a
gauge invariant amplitude by analyzing a particular reaction in detail; this
is precisely what has frequently been done in $Z$ physics at LEP I.  In this
paper we present a more general method to obtain a gauge invariant amplitude
based on the analytical properties of the amplitude.

This procedure to treat unstable particles has been given a long time ago by
M.~Veltman \cite{VeltmanUnstable}.  Recently it has been applied to the
production and decay of a single uncharged particle, the $Z$ boson, to first
order, in Refs \cite{Zphysics,Stuart1,HVeltmanUnstable}.  The method centers
on the isolation
of the position and residue of the unstable particle pole from a fixed-order
calculation.  We generalize it to the treatment of multiple, charged unstable
particles.  Both extensions bring new effects at the one loop level.
A side effect worth mentioning is that the non-resonant terms, which are not
enhanced by the resonant propagator, are separated in a gauge invariant way.
This means that they can be computed to one order less than the resonant
terms.  This greatly economizes on the amount of work necessary.
The reaction we have in mind is of course $W$ pair production at LEP II; in
this paper we present the general formalism.

The layout of this article is as follows.  First we review some alternative
methods which have been proposed to treat unstable particles in perturbation
theory.  Next we separate the diagrams for a given reaction into three groups:
factorizable resonant, non-factorizable resonant and non-resonant.  The first
group is analyzed to all orders for a single uncharged particle in section
\ref{sec:fact}.  A list of subtleties is discussed in section \ref{sec:fine};
this includes the effects of a charged particle.   Next we analyze the
non-factorizable diagrams.  Finally we recapitulate the gauge invariance
arguments, and give a recipe for tree level and one loop calculations in this
scheme.  The appendices contain various proofs needed in the text.


\section{The Finite Width and Kinematics}
\label{sec:kin}

The consistent treatment of unstable particles cannot be implemented on the
level of the Lagrangian.  A proposal using this idea \cite{Stuart2} is to add
and subtract a piece proportional to the width in the mass term of the
original Lagrangian: $m^2\phi^2 \to (m^2 -im\Gamma)\phi^2 + im\Gamma\phi^2$.
The first term now gives a complex mass $M^2 = m^2-im\Gamma$ in the
propagator, the second one is taken as a perturbatively treated two-particle
interaction.  In the standard model this leads to a complex weak mixing
angle $\sin^2\theta_W = 1-M_W^2/M_Z^2$, as $M_W^2$ and $M_Z^2$ are both
complex.

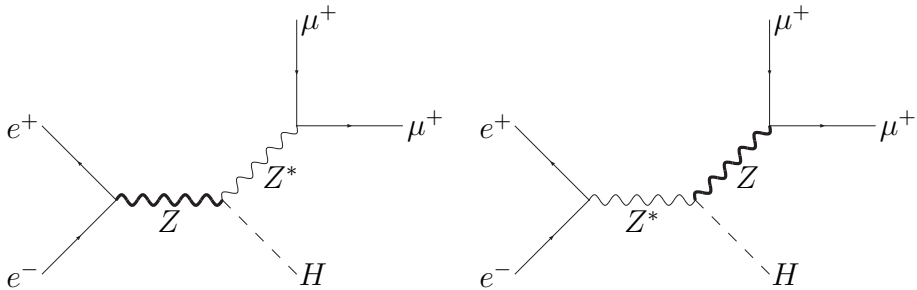
\begin{figure}[htb]
\unitlength .4bp
\setscale{.4}
\centerline{
\begin{picture}(400,270)(0,0)
\ArrowLine(30,30)(100,100)
\ArrowLine(100,100)(30,170)
\Vertex(100,100){.8}
{
\def\axowidth{3 }
\Photon(100,100)(200,100){5}{5}
}
\Vertex(200,100){.8}
\DashLine(200,100)(270,30){10}
\Photon(200,100)(270,170){5}{5}
\Vertex(270,170){.8}
\ArrowLine(270,270)(270,170)
\ArrowLine(270,170)(370,170)
\put(28,30){\makebox(0,0)[r]{\large$e^-$}}
\put(28,170){\makebox(0,0)[r]{\large$e^+$}}
\put(150,90){\makebox(0,0)[t]{\large$Z$}}
\put(272,30){\makebox(0,0)[l]{\large$H$}}
\put(238,132){\makebox(0,0)[tl]{\large$Z^*$}}
\put(274,270){\makebox(0,0)[l]{\large$\mu^+$}}
\put(374,170){\makebox(0,0)[l]{\large$\mu^+$}}
\end{picture}
\hss
\begin{picture}(400,270)(0,0)
\ArrowLine(30,30)(100,100)
\ArrowLine(100,100)(30,170)
\Vertex(100,100){.8}
\Photon(100,100)(200,100){5}{5}
\Vertex(200,100){.8}
\DashLine(200,100)(270,30){10}
{
\def\axowidth{3 }
\Photon(200,100)(270,170){5}{5}
}
\Vertex(270,170){.8}
\ArrowLine(270,270)(270,170)
\ArrowLine(270,170)(370,170)
\put(28,30){\makebox(0,0)[r]{\large$e^-$}}
\put(28,170){\makebox(0,0)[r]{\large$e^+$}}
\put(150,90){\makebox(0,0)[t]{\large$Z^*$}}
\put(272,30){\makebox(0,0)[l]{\large$H$}}
\put(238,132){\makebox(0,0)[tl]{\large$Z$}}
\put(274,270){\makebox(0,0)[l]{\large$\mu^+$}}
\put(374,170){\makebox(0,0)[l]{\large$\mu^+$}}
\end{picture}
}
\caption[]{The Bjorken process at $\sqrt{s}\approx m_Z$ (left) and
$\sqrt{s}\gsim m_Z+m_H$ (left).  $Z^*$ denotes an off-shell $Z$ boson; the
thick line indicates a propagator which needs to be resummed.}
\label{fig:Bjorken}
\end{figure}

The major drawback is that this Lagrangian gives rise to unwanted effects in
case the unstable particles are {\em not\/} near their mass shell.  One
example would be deep inelastic scattering, where the t-channel $W$ boson
obtains an on-shell width $\Gamma_W$, in spite of the fact that the self
energy is purely real for spacelike $p^2$.  Another example is the Bjorken
mechanism for Higgs production: $e^+e^- \to H\mu^+\mu^-$ (see
Fig.~\ref{fig:Bjorken}).  The two diagrams may look similar, but have to be
treated differently.  At $\sqrt{s}\approx m_Z$ (LEP I) it is necessary to
resum the first $Z$ boson, but not the second one; at $\sqrt{s}\gsim m_H+m_Z$
the second one can be on-shell.  Using the on-shell width in the first $Z$
boson propagator is in fact wrong: as the self energy scales roughly with $s$
one should use a width which is four times larger than the on-shell width at
LEP II energies.  This error can be corrected in this scheme, but this
requires a one loop calculation (introducing errors of $\order(\alpha^2)$).

The treatment of the unstable particle propagator depends therefore critically
on the kinematical configuration.  We will refer to a particle as unstable
only when it can kinematically be on its mass shell.  In this case we are
forced to resum its propagator.  Otherwise the particle can and should be
treated perturbatively.

The basic idea of the method used here is to consider the analytical
properties of the transition amplitude.  Neglecting for the moment all further
details and qualifiers, the amplitude containing one unstable particle has the
form
\begin{equation}
\label{eq:ampform}
    {\cal A} = \frac{w}{p^2-M^2} + n(p^2) \;,
\end{equation}
in which $p$ is the momentum flowing through the unstable particle propagator,
$M^2$ the (complex) pole position, $w$ the residue at this pole and $n(p^2)$
the non-resonant remainder.  The pole position, the residue at the pole and
the remainder are gauge invariant quantities.  We will show how these can be
computed systematically and show this gauge invariance explicitly.  Initially
we will discuss this for a single particle.  The generalization to more
unstable particles is given afterwards.


\section{Classes of Diagrams}


In order to simplify the discussion we will divide the diagrams --- of a
perturbative expansion of the amplitude to all orders --- into three classes.
This separation is not gauge invariant; we use it to extract the gauge
invariant contributions to the residue at the pole in Eq.~(\ref{eq:ampform})
more easily.


\begin{figure}[htb]
\centerline{
\setscale{.4}
\unitlength .4bp
\begin{picture}(400,160)(0,20)
\ArrowLine(20,20)(100,100)
\ArrowLine(20,180)(100,100)
\ArrowLine(100,100)(180,20)
\ArrowLine(100,100)(180,180)
{
\def\axowidth{3 }
\Photon(100,100)(400,100){5}{15}
}
\ArrowLine(400,100)(500,100)
\ArrowLine(400,100)(485,150)
\ArrowLine(400,100)(485, 50)
\GCirc(100,100){25}{.8}
\GCirc(250,100){25}{.8}
\GCirc(400,100){25}{.8}
\put(100,0){\makebox(0,0)[b]{\small production}}
\put(250,0){\makebox(0,0)[b]{\small propagator}}
\put(400,0){\makebox(0,0)[b]{\small decay}}
\end{picture}
}
\caption{The structure of the factorizable resonant diagrams}
\label{fig:fact}
\end{figure}
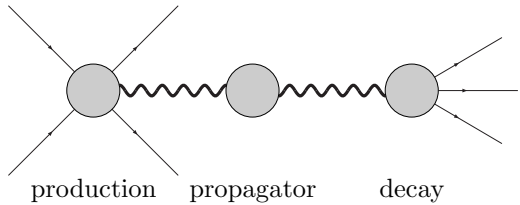

The first class of diagrams we call the factorizable resonant diagrams (see
Fig.~\ref{fig:fact}).  It comprises all diagrams that retain the original
unstable particle propagator $1/(p^2-m^2)$ outside the loops.  These are the
diagrams appearing in the narrow width approximation.  The radiative
corrections occur separately in the production, the decay and the propagator,
but do not connect production and decay (or, in the case of more resonances,
the decays of different resonances).  The contribution to the amplitude can be
written as
\begin{eqnarray}
\label{eq:Ainffact}
    {\cal A}^\infty_{\rm fact} & = &
\frac{W(p^2,\ldots)}{p^2-m^2}\sum_{n=0}^\infty
        \left(\frac{\Pi(p^2)}{p^2-m^2}\right)^n
\;,
\end{eqnarray}
where $m$ denotes the renormalized mass, which is real and finite, but need
not be the on-shell mass (which is defined as the real part of the pole
position).  The corrections to production and decay are contained in $W$, and
$\Pi(p^2)$ is the one particle irreducible self energy.  This part of the
amplitude is infrared divergent.  The structure of the soft bremsstrahlung
diagrams which have to be added to cancel this divergence is identical (after
integrating out the extra soft photons).

The ellipsis in the argument of $W$ stands for other variables which determine
the kinematical configuration and hence the the amplitude.  Examples are the
angles of the resonance decay products and the momentum squared of particles
produced far off their mass shell (for instance the second $Z$ boson in the
Bjorken mechanism for Higgs production at $\sqrt{s} \approx m_Z$).  It is
important that the integration boundaries of these variables are independent
of $p^2$, unlike, e.g., Mandelstam invariants.   This would introduce extra
terms in the expansions used below.  On the other hand, angles in the CMS are
obviously a good choice.  If this does not suffice to eliminate all variables
with $p^2$ dependent integration boundaries (like the momenta squared
mentioned before) one has to introduce explicit mappings from variables
$0<x_i<1$ to these $p^2$ dependent variables.  The angles and mappings are not
unique; one can for instance change the frame in which the angles are defined.
 Different choices will change the answer in a fixed order calculation, but
the difference will be of higher order.  An explicit proof for the
$\order(\alpha)$ case can be found in appendix \ref{ap:frame}.

This dependence on other variables introduces problems when the first argument
of $W$ is such that the kinematical configuration becomes unphysical. This
happens when $p^2$ is below the threshold for production or decay of the
unstable particle.  We choose to define $W(p^2)$ to be zero in this region.
However, this will cause problems near, and especially below threshold.


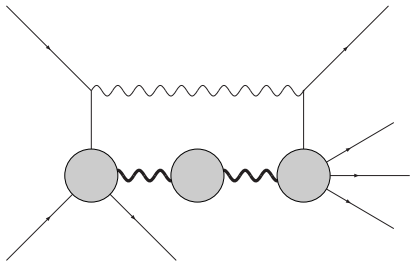
\begin{figure}[htb]
\centerline{
\setscale{.4}
\unitlength .4bp
\begin{picture}(400,240)(0,20)
\ArrowLine(20,20)(100,100)
\Line(100,180)(100,100)
\ArrowLine(20,260)(100,180)
\ArrowLine(100,100)(180,20)
\Vertex(3100,180){.8}
{
\def\axowidth{3 }
\Photon(100,100)(300,100){5}{10}
}
\ArrowLine(300,100)(400,100)
\ArrowLine(300,100)(385,150)
\ArrowLine(300,100)(385, 50)
\Line(300,100)(300,180)
\ArrowLine(300,180)(380,260)
\Vertex(300,180){.8}
\Photon(100,180)(300,180){5}{10}
\GCirc(100,100){25}{.8}
\GCirc(200,100){25}{.8}
\GCirc(300,100){25}{.8}
\end{picture}
}
\caption{The structure of the non-factorizable resonant diagrams}
\label{fig:nonfact}
\end{figure}

There is a second class of diagrams which also diverges linearly for $p^2\to
m^2$.  This group consists of all infrared divergent (photonic and gluonic)
corrections to the lowest order process which span production and decay (or
the decay of different unstable particles).  A prototype for this kind of
diagram is given in Fig.~\ref{fig:nonfact}.  The linear divergence for $p^2\to
m^2$ is related to the infrared divergence in that both occur in the limit
that the photon momentum in the loop $Q\to0$.  However, whereas the infrared
divergence is logarithmic, the on-shell divergence is linear (or higher for
more unstable particles).  In both cases the divergence cancels against the
corresponding soft Bremsstrahlung graphs (which have to be treated analogously
to the virtual graphs), but only after integrating in the soft-photon
approximation the photon up to an energy much larger than the width.


Thirdly, all other diagrams are non-resonant, i.e., they do not diverge in the
limit $p^2\to m^2$.  Examples of these are diagrams leading to the same final
state without the resonant propagator, and all non-infrared divergent graphs
connecting production and decay (or different decays of unstable particles).
These graphs can not and should not be resummed.

In this classification scheme we assume that there are no two channels
involving different resonances leading to the same final state.  An example
violating this is $e^+e^- \to \mu^+\mu^-\nu\bar{\nu}$, which can proceed via
two $W$ bosons or two $Z$ bosons.  These two contributions peak in different
regions of phase space, but there is an overlap.  We do not address this
issue in this paper.



\section{The Factorizable Resonant Diagrams}
\label{sec:fact}

In this section we discuss how to find the position of the pole and the
contribution to its residue from the factorizable diagrams.  Both these
quantities are gauge invariant.  Outside the resonance region $p^2 \approx
m^2$, Eq.~(\ref{eq:Ainffact}) can just be expanded perturbatively.  However,
near the pole $p^2=m^2$ a fixed-order perturbative calculation is useless;
instead we will have to resum part of the higher order self energy
contributions.  This introduces problems in the case of charged unstable
particles.  In this section we assume that the resonance is
neutral\footnote{that is, neutral with respect to any unbroken gauge group.};
the extension to charged particles will be given in the next section.

After summing the geometrical series we isolate the pole structure as
\cite{Stuart1,HVeltmanUnstable}
\begin{eqnarray}
   \frac{W(p^2)}{p^2-m^2-\Pi(p^2)}
        & = & \left[\frac{W(p^2)}{p^2-m^2-\Pi(p^2)}
                - \frac{W(M^2)}{p^2-M^2}\frac{1}{1-\Pi'(M^2)} \right]
            + \frac{W(M^2)}{p^2-M^2}\frac{1}{1-\Pi'(M^2)}
\nonumber\\
        & \stackrel{\rm def}{=} & n(p^2) + \frac{w}{p^2-M^2} \;,
\label{eq:Mexpansion}
\end{eqnarray}
with the pole position defined by $M^2 - m^2 - \Pi(M^2) = 0$.  (If $m^2$
is the on-shell mass $M^2 = m^2 - im\Gamma$.  This differs from the standard
on-shell renormalization scheme by the fact that we use $\Pi(M^2)$ in the
counterterm, rather than $\Pi(m^2)$; the difference is of $\order(\alpha^2)$.)
 Here, $w$ and $\Pi(M^2)$ are gauge invariant.  However, the quantity
$W(M^2,\ldots)$ is not yet well-defined because of the dependence on the other
arguments: if $p^2$ is off the real axis the angles also have to be
taken complex.  However, we do not need to evaluate $W(M^2,\ldots)$, as the
residue can be read off from an alternative expansion of the first class of
diagrams around the real mass $m^2$.  This gives
\begin{equation}
    \frac{W(p^2)}{p^2-m^2}\sum_{n=0}^\infty
            \left(\frac{\Pi(p^2)}{p^2-m^2}\right)^n
        = \bar{N}(p^2) + \frac{W_{-1}}{p^2-m^2}
            + \sum_{n=2}^\infty\frac{W_{-n}}{(p^2-m^2)^n} \;.
\label{eq:mexpansion}
\end{equation}
The coefficients can be found as a series in $\alpha\sim\Pi$ by a
straightforward expansion; the first two terms are given by
\begin{eqnarray}
    W_{-1} & = & W(m^2) + \frac{d}{dp^2}\bigl[W(p^2)\Pi(p^2)\bigr]_{p^2=m^2} +
         \frac{1}{2}\frac{d^2}{d(p^2)^2}\bigl[W(p^2)\Pi^2(p^2)\bigr]_{p^2=m^2}
+
         \cdots \;.
\label{eq:resonant}
\\
    \bar{N}(p^2) & = & \frac{W(p^2)-W(m^2)}{p^2-m^2}
\nonumber\\&&\mbox{}
      + \frac{W(p^2)\Pi(p^2) - W(m^2)\Pi(m^2)
      -
(p^2-m^2)\frac{d}{dp^2}\bigl[W(p^2)\Pi(p^2)\bigr]_{p^2=m^2}}{(p^2-m^2)^2}
        + \cdots
\label{eq:nonreson}
\end{eqnarray}
(In the last term we have chosen not to continue the expansion in positive
powers of $(p^2-m^2)$, obtaining differences rather than derivatives.)
As was shown in Refs \cite{Stuart1,HVeltmanUnstable}, the two expressions
(\ref{eq:Mexpansion}) and (\ref{eq:mexpansion}) correspond rather nicely in
low orders in $\alpha$ and $(p^2-m^2)$.  In fact, it turns out that
\begin{eqnarray}
\label{eq:w=W}
        w & = & W_{-1}
\\
\label{eq:n=N}
        n(p^2) & = & \bar{N}(p^2)
\end{eqnarray}
to {\em all\/} orders, and thus the residue at the pole $1/(p^2-M^2)$ and the
non-resonant remainder can be obtained directly from a perturbative off-shell
calculation.   A proof is given in appendix \ref{ap:proofs}.  To the best of
our knowledge, these relations have not been stated in full generality before.
They permit us to find the residue at the pole and the non-resonant parts of
the resummed expression (\ref{eq:Mexpansion}) from the unresummed one
(\ref{eq:mexpansion}).  The latter is a straightforward perturbation
expansion.

Let us discuss the terms needed for a one loop calculation and their
interpretation.  The first order result is given by
\begin{eqnarray}
    {\cal A}^{(0)} & = & \frac{W^{(0)}(m^2)}{p^2-M^2} \\
    M^2 & = & m^2 + \Pi^{(1)}(M^2)
\;,
\end{eqnarray}
integrated over the full phase space.  We have dropped the subscript
factorizable, as the non-factorizable and non-resonant terms do not give a
contribution in this order.  The superscripts indicate to which order the
different parts are computed: we need the one loop self energy at the pole (in
fact only the imaginary part) and the on-shell amplitude at tree level.
The limit $\Gamma = i\Pi(M^2)/m \to 0$ in the propagator and phase space as
well gives the narrow width approximation.

In the next approximation one has to use the two loop corrected on-shell self
energy $\Pi^{(1+2)}(M^2)$ (which gives the physical width) in the definition
of the pole position in the previous term.  Besides this one finds the
following contributions.
\begin{itemize}
\item The $\order(\Gamma)$ corrections
\begin{equation}
    {\cal A}^{(\Gamma)} = \bar{N}^{(0)}(p^2) =
        \frac{W^{(0)}(p^2) - W^{(0)}(m^2)}{p^2-m^2}
\;.
\end{equation}
These corrections (and the non-resonant contributions) have been
studied in Ref.\ \cite{OGamma}.
\item The one loop on-shell correction with resonant propagator
\begin{equation}
\label{eq:oneloop}
    {\cal A}^{(1)}_{\mbox{fact}} = \frac{W^{(1)}(m^2)}{p^2-M^2}
\;.
\end{equation}
\item The derivative term
\begin{equation}
    {\cal A}^{(Z)} = \frac{W^{(0)}(m^2)\Pi^{(1)}{}'(m^2)}{p^2-M^2}
\;.
\end{equation}
In the limit $\Gamma\to0$ this term compensates for the fact that, in the
narrow width approximation, one has twice a field renormalization of the
resonant field $(Z_\phi)^{1/2} = 1/\sqrt{1-\Pi'(m^2)}$. It also implements in
a perturbative way the $p^2$-dependence of the self energy, which is
included in most descriptions of the $Z$ pole.
\item The term
\begin{equation}
    {\cal A}^{(W')} = \frac{W^{(0)}{}'(m^2)\Pi^{(1)}(m^2)}{p^2-M^2}
\;.
\end{equation}
(which vanishes in the limit $\Gamma\to0$)
compensates at this order in $\alpha$ the dependence on the choice of the
other kinematical variables mentioned in section \ref{sec:kin}.
A proof is given in appendix \ref{ap:frame}.
\end{itemize}

In the case that there are more unstable particles which can simultaneously be
on their mass shell, the above resummation can be performed for each pole
separately.  The resulting series provides a natural way to isolate
the relevant contributions.  In the case of two unstable particles these
are the doubly resonant terms (both unstable particles on their mass shell)
for a tree level calculation, and the one loop doubly resonant and tree level
singly resonant terms for the $\order(\alpha)$ corrections.  The non-resonant
terms only occur at $\order(\alpha^2)$ (with the possibility of large logs).


\section{The Fine Print}
\label{sec:fine}


The formulae in section \ref{sec:fact} look concise, but several subtleties
have not been addressed; we will treat them in this section.  We first
investigate the threshold behavior and show that the method developed here
breaks down in this region.  Next we consider the definitions of $W$ and $\Pi$
in greater detail, and treat the phenomena of mixing and unphysical particles.
Finally we expose the additional problems that occur when there are charged
unstable particles.  The difficulties with the proper definition of the
kinematical variables have already been addressed in section \ref{sec:kin} and
appendix \ref{ap:frame}.


\subsection{Threshold problems}

The expansions (\ref{eq:Mexpansion}) and (\ref{eq:mexpansion}) introduce
difficulties near the threshold for the production or decay
of the unstable particle.
In the case of the $Z$ peak this is not a problem, as the CMS energy squared
$s$ coincides with $p^2$.  However, this is not the case when more particles
are produced, for instance $e^+e^-\to H\mu^+\mu^-$ at $s\approx (m_H+m_Z)^2$
via the Bjorken mechanism.  The leading term in the expansion now is $W(m_Z^2,
\ldots)$; but if $s < (m_H+m_Z)^2$, the point $p^2=m_Z^2$ is outside the
physical region.  The angles between the Higgs and Z bosons would have to be
complex to satisfy energy-momentum conservation.  As stated before, we choose
to define $W(p^2,\ldots) = 0$ for unphysical $p^2$ to avoid these problems.
This, however, effectively annuls the whole resummation scheme and one finds
back the original $1/(p^2-m^2)$ divergence in $\bar{N}(p^2)$ as the threshold
is approached from below.  Far below threshold no resummation is needed and no
problems occur.  The bad behavior will also manifest itself when approaching
the threshold from above due to the occurrence of terms
$\Gamma/(\mbox{distance from threshold})$.  Therefore, with the choice
$W(p^2)=0$ for unphysical $p^2$ one cannot use the expansion very close to
threshold. One should thus either not use the separation in resonant and
non-resonant contributions (which is possible for the Bjorken process and has
been done until now \cite{Fleischer&JegerlehnerZH,WuerzburgZH,KniehlZH}), or
attempt an expansion in other variables or an non-relativistic approach
\cite{Kuhntop}.


\subsection{The definition of the self energy $\Pi$}

The split between the self energies $\Pi(p^2)$, which are resummed in Eq.\
(\ref{eq:Mexpansion}), and the corrections in $W(p^2)$ to production or decay,
is not unique.  Pieces of $\Pi(p^2)$ that vanish at the pole position, and
thus do not contribute to the physical width, can also be assigned to the
perturbatively treated part $W(p^2)$.  In fact, one can choose to treat {\em
all\/} $p^2$ dependence in the self energy perturbatively which implies adding
terms proportional to $\Pi(p^2)-\Pi(M^2)$ to $W(p^2)$.  In this way one can
achieve $\Pi'(m^2)=0$ and the equivalence of the two series
(\ref{eq:Mexpansion}) and (\ref{eq:mexpansion}) is trivial.

In the case of the $Z$ and $W$ bosons it is customary to resum the
$p^2$-dependent fermionic self energy.  This is a gauge invariant quantity,
even off-shell.  The bosonic self energy does not contribute on-shell and is
calculated perturbatively.  This perturbative treatment is required by the
gauge cancellations with vertex corrections and boxes.  The scheme given here
is different, in that we only resum the {\em physical\/} width, and also treat
the $p^2$ dependence of the fermionic self energy perturbatively.  The
difference is always of higher order in $\alpha$.


\subsection{Mixing and unphysical particles}

At one loop and higher there are graphs converting physical particles into
each other, such as the photon and $Z$ boson (see for instance
\cite{Stuart1,HVeltmanUnstable,FleischerJegerlehnerHiggs1,DennerHabilitation}).
The $2\times2$ matrix for the self energies can easily be diagonalized; the
resulting eigenvectors define the $n$ loop photon and $Z$ propagators. The $Z$
propagator can then be resummed as before.

In a general gauge there also is a plethora of unphysical particles included
in the standard model Lagrangian: Fadeev-Popov ghosts and the unphysical Higgs
bosons.  These have masses which depend on the gauge parameter $\xi$, and it
is not clear how these should be resummed, although various suggestions have
been made \cite{Pilaftsis}.  As we isolate the pole structure from a fixed
order off-shell calculation with real masses via Eqs (\ref{eq:w=W}) and
(\ref{eq:n=N}) all gauge dependence has already disappeared before the
resummation is performed.  Only the physical width, which cannot depend on the
gauge parameter, is resummed and one never encounters these unphysical
particles in the parts taken from the higher order graphs.


\subsection{Charged particles.}
\label{subsec:charged}

\begin{figure}[htb]
\begin{center}
\begin{picture}(150,100)(0,0)
\Vertex(30,50){2}
{\def\axowidth{1.2 }
\Photon(0,50)(30,50){2}{3}
\Photon(30,50)(77.6,77){2}{5.5}
}
\Photon(30,50)(77.6,23){-2}{5.5}
\ArrowLine(99.3,90)(77.6,77)
\ArrowLine(77.6,77)(77.6,23)
\ArrowLine(77.6,23)(99.3,10)
\Vertex(77.6,77){2}
\Vertex(77.6,23){2}
\put(-2,50){\makebox(0,0)[r]{$W^-$}}
\put(80,50){\makebox(0,0)[l]{$\mu$}}
\put(52,66){\makebox(0,0)[br]{$W$}}
\put(50,31){\makebox(0,0)[tr]{$\gamma$}}
\put(102,90){\makebox(0,0)[l]{$\bar{\nu}_\mu$}}
\put(102,10){\makebox(0,0)[l]{$\mu^-$}}
\put(50,0){\makebox(0,0)[l]{I}}
\put(60,50){\makebox(0,0){$Q$}}
\LongArrowArc(60,50)(10,180,300)
\end{picture}
\begin{picture}(110,100)(0,0)
\Photon(0,50)(30,50){2}{3}
\Vertex(30,50){2}
{\def\axowidth{1.2 }
\Photon(30.3,50)(99.6,90){2}{8}
\Photon(30.3,50)(99.6,10){2}{8}
}
\PhotonArc(30,50)(55,-30,+30){2}{5}
\Vertex(77.6,77.5){2}
\Vertex(77.6,22.5){2}
\put(-2,50){\makebox(0,0)[r]{$\gamma^*,Z^*$}}
\put(88,50){\makebox(0,0)[l]{$\gamma$}}
\put(52,66){\makebox(0,0)[br]{$W$}}
\put(52,33){\makebox(0,0)[tr]{$W$}}
\put(102,90){\makebox(0,0)[l]{$W^+$}}
\put(102,10){\makebox(0,0)[l]{$W^-$}}
\put(65,50){\makebox(0,0){$Q$}}
\LongArrowArc(65,50)(10,180,300)
\put(50,0){\makebox(0,0)[l]{II}}
\end{picture}
\end{center}
\caption{The two on-shell divergent scalar three point functions with
physical examples}
\label{fig:Cs}
\end{figure}
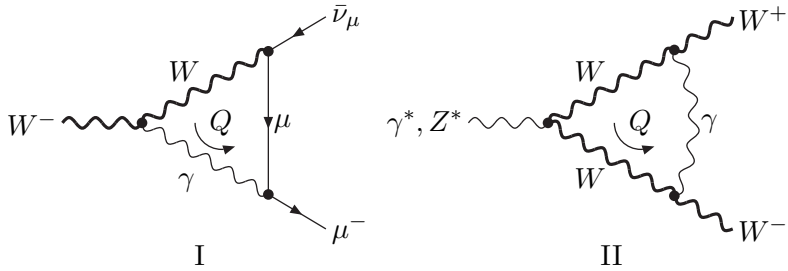

When the unstable particle is (electrically or color) charged, a complication
arises from logarithmically divergent on-shell graphs.  These are graphs that
contain terms proportional to $\log(p^2-m^2)$ because the massless particle
couples to the resonance.  Examples of one loop graphs of this type are given
in Fig.~\ref{fig:Cs}; these graphs would be infrared divergent if the
unstable particle would be on its mass shell.  This means that $W^{(1)}(m^2)$
is not well-defined (not regulating this divergence with a photon mass),
invalidating the expansion (\ref{eq:mexpansion}).  One could argue that
these on-shell divergent logarithms should cancel against the corresponding
soft bremsstrahlung graphs, as $p^2-m^2\ne0$ is just another infrared
regulator, on which the physical result should not depend.  Unfortunately this
only holds in the limit $E_\gamma<\omega\ll$ all other scales (to validate the
soft-photon approximation), but $\omega\gg\Gamma$ (so that $\Gamma$ is only an
infrared regulator).  In case the width is non-negligible these two
requirements are clearly incompatible.

We can still use the expansion techniques by defining the
function $W^{(1)}(m^2)$ in the representation of the amplitude as a sum of
fundamental loop integrals times coefficient functions.  Fundamental integrals
are a set of linearly independent integrals in which all tensor integrals can
be expressed.  In the case of one loop corrections, these are the scalar one
loop integrals with non-zero kinematical determinants, but for higher loops
there are also fundamental tensor integrals which cannot be algebraically
reduced to scalar integrals.  A proof that in a one loop amplitude the scalar
integrals are the fundamental integrals is given in appendix
\ref{ap:scalargauge}.  This independence implies that the coefficient
functions of physical fundamental integrals (integrals that depend only on
physical masses) are gauge invariant, and the coefficients of unphysical
fundamental integrals (those containing a gauge parameter $\xi$) must vanish.
Now, the $\log(p^2-m^2)$ terms only occur in the scalar functions, as the
coefficient function is a rational function.  We are therefore free to include
finite width effects in the scalar functions without concern about gauge
invariance.

An obvious approach is to apply the expansion (\ref{eq:Mexpansion}) to the
unstable particle propagators {\em before\/} the integration over the loop
momentum $Q$ when evaluating these scalar functions. Up to $\order(\alpha)$,
the complicated non-resonant terms in Eq.~\ref{eq:mexpansion} do not
contribute.  In the simplest case, the vertex corrections, one just obtains
the scalar three point function with complex masses. The evaluation of this
function was already described in Ref.\ \cite{tHooft&Veltman}.  A numerically
stable implementation is given in Refs \cite{NewAlgorithms,FFguide}.  Note
that one should only use a complex mass in propagators which  are resonant in
the infrared limit $Q\to0$, as in the other propagators the expansion of the
self energy around $p^2=M^2$ does not describe its behavior well in the
dominant region.  The extension to higher order graphs would be much more
involved.

The higher point functions can now also easily be evaluated.  One just
subtracts from the infrared divergent $n$ point function with real masses the
corresponding infrared divergent three point function (suitably
scaled\footnote{One can either multiply by the non-infrared propagators, or the
ratio of overall determinants occurring in the scalar functions.}), and
adds the same three point function with complex masses.
For the four point function this is just the first term of the Taylor
expansion proposed in Ref.\ \cite{cD0}.\footnote{This article contains an
error, in that to obtain a convergent Taylor series it is not sufficient to
{subtract\/} the non-analytic parts, but one has to {replace\/} these by the
off shell expressions.}  This way also the large cancellations between three
and
four point functions near the edge of phase space ($|\cos\theta_{eW}|
\approx1$ in the case of $W$ pair production) are unaffected by the width.
Note that {\em on} the edge the four point function reduces to a sum of three
point functions.

The prescription followed for the scalar integrals must be matched, of course,
by the prescription used in the soft bremsstrahlung integrals; the lower end
of which should be the pole at $Q^2=0$ of the scalar functions.

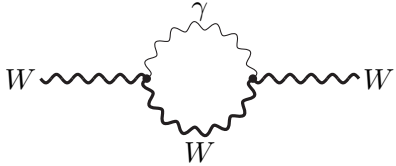
\begin{figure}[htb]
\setscale{.8}
\unitlength .8bp
\begin{center}
\begin{picture}(150,60)(0,10)
{\def\axowidth{1.5 }
\Photon(  0,40)( 50,40){-2}{5}
\PhotonArc(75,40)(25,-180,-360){2}{8}
\Photon(100,40)(150,40){-2}{5}
}
\PhotonArc(75,40)(25,0,-180){2}{8}
\Vertex( 50,40){2}
\Vertex(100,40){2}
\put( -2,40){\makebox(0,0)[r]{$W$}}
\put(152,40){\makebox(0,0)[l]{$W$}}
\put( 75,10){\makebox(0,0)[t]{$W$}}
\put( 75,70){\makebox(0,0)[b]{$\gamma$}}
\end{picture}
\end{center}
\caption{Example of a diagram which is not analytic at $p^2=m^2$: the one
loop photonic contribution to the self energy of a $W$ boson.}
\label{fig:Wg}
\end{figure}

There is one more graph which spoils the expansions; this is the self energy
graph of Fig.~\ref{fig:Wg}, which also is non-analytic at $p^2=m^2$.  The
structure of this diagram is given by
\begin{equation}
\label{eq:Wgstruct}
    {\cal A}^\gamma_{\rm off} = \frac{g_1(p^2) + g_2(p^2)
        \displaystyle\frac{B_0(p^2,m^2,0) - B_0(m^2,m^2,0)}{p^2-m^2}}{p^2-m^2}
    \;,
\end{equation}
with the $g_i(p^2)$ non-singular functions, $B_0$ the one loop scalar two
point function and
\begin{equation}
\label{eq:diffB0}
    \frac{B_0(p^2,m^2,0) - B_0(m^2,m^2,0)}{p^2-m^2} = \frac{-1}{p^2}
        \log\left(\frac{p^2-m^2}{m^2}\right)
    \;.
\end{equation}
A resummation of the propagators before integration therefore gives a complex
mass in the one remaining overall propagator in Eq.\ (\ref{eq:Wgstruct}) and
in the internal propagator, which just means replacing $m^2\to M^2$ everywhere
in Eq.\ (\ref{eq:diffB0}).  This also shifts the subtraction point from $m^2$
to $M^2$, as required in section \ref{sec:fact}.



\subsection{Further refinements}

As mentioned above, the details of the evaluation of the scalar functions do
not influence the gauge invariance of the final result.  For any non-singular
approximation of the scalar function we have a gauge invariant residue.
We will mention some variants of the method just described.

\begin{itemize}
\item One can use complex masses in the divergent logarithms only, rather than
when resonant in the scalar function.   The difference is formally of order
$\alpha\Gamma/\mu$ only, with $\mu$ some scale.  However, this scale may be
related to the distance from threshold rather than the mass of the resonance,
so in that case one may miss some important threshold effects.  An example is
the Coulomb singularity, which would not be smeared by the width.

\item One can use a complex mass everywhere in the scalar functions.  As
mentioned above, this introduces errors, but these are of one order higher in
$\alpha$ and thus irrelevant.

\item There are terms $(p^2-m^2)\log(p^2-m^2)$ associated with the
non-infrared radiation of a photon off a charged unstable particle.  As these
also originate at $Q\approx0$ their resummation to $(p^2-M^2)\log(p^2-M^2)$
rather than putting them to zero should improve the resonant behavior.  An
example of a diagram in which this occurs is given in Fig.~\ref{fig:thres0}.

\begin{figure}[htb]
\begin{center}
\begin{picture}(80,70)(20,0)
\Photon(20, 0)( 50,10){2}{3}
\Photon(20,70)( 50,60){2}{3}
{\def\axowidth{1.5 }
\Photon(50,10)(100, 0){2}{5}
\Photon(50,60)(100,70){2}{5}
}
\Photon(50,60)( 50,10){2}{5}
\Vertex(50,10){2}
\Vertex(50,60){2}
\Vertex(50,35){2}
\Vertex(74.51,64.90){2}
\PhotonArc(50,60)(25,-90,11.31){2}{5}
\put(18, 0){\makebox(0,0)[r]{$\gamma$}}
\put(18,70){\makebox(0,0)[r]{$\gamma$}}
\put(104, 0){\makebox(0,0)[l]{$W^+$}}
\put(104,70){\makebox(0,0)[l]{$W^-$}}
\put( 46,30){\makebox(0,0)[r]{$W$}}
\put( 70,40){\makebox(0,0)[tl]{$\gamma$}}
\end{picture}
\end{center}
\caption{Example of a subdiagram with a photonic threshold.  The
(predominantly)
t-channel $W$ propagators are of course not resummed.}
\label{fig:thres0}
\end{figure}
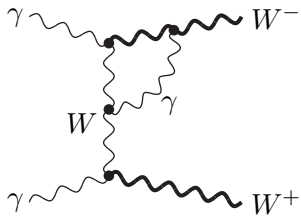

\item Similarly, at the threshold for on-shell production of two particles
there is a term
\begin{equation}
    \frac{\sqrt{-\delta}}{p^2}
    \log\left(\frac{m_1^2+m_2^2-p^2-2\sqrt{-\delta}}{2m_1m_2}\right)
    \;,
\end{equation}
with $\delta = m_1^2 m_2^2 - (p^2-m_1^2-m_2^2)^2/4$ related to the K\"all\'en
function.  At this threshold, the scalar function is also dominated by
$Q\approx0$, so including a complex mass for these particles should give a
better approximation.  This happens for instance in the diagrams of
Fig.~\ref{fig:Cs} with the $\gamma$ replaced by a $Z$ and $q^2 \approx
(m_1+m_2)^2$.

\item It is not necessary to put $p^2=m^2$ in the non-divergent parts of the
evaluation of the scalar functions; in fact, leaving the momenta off shell
(but the masses real) will also improve the threshold behavior.
\end{itemize}



\section{The non-factorizable resonant diagrams}
\label{sec:nonfact}


We now turn to the non-fac\-tor\-iz\-able resonant diagrams.  They are absent
at tree level.  We will discuss them in three steps.  The first one is to
separate the resonant contribution of these diagrams from the non-resonant
part.  The non-resonant part can easily be evaluated, but the on-shell
divergence of the resonant terms will have to be resummed.  After showing how
this can be done for one loop integrals we note the explicit structure of the
divergent terms in the integrals needed for $e^+e^-\to4\mbox{ fermions}$.
Finally, we comment on the cancellations between these diagrams and the
corresponding bremsstrahlung integrals.


\subsection{Resummation}

Before discussing the structure of the non-factorizable diagrams it is useful
to recapitulate the resummation of the factorizable diagrams.
The separation between resonant and non-resonant terms in the factorizable
diagrams was made implicitly in section \ref{sec:fact} by writing the
corrections to production and decay as $W(p^2) = W(m^2) + [W(p^2)-W(m^2)]$,
where only the propagator multiplying the on-shell contribution was resummed.
This had the advantage that the second term is needed to one order less, as
it is suppressed by a power of $\Gamma\sim\alpha$.  The first term is gauge
invariant (this will be shown in section \ref{sec:gauge}).
This scheme breaks down when $W(m^2)$ contains logarithmic divergences (a
charged unstable particle).  The solution is to write the amplitude as a sum
of fundamental loop integrals, the coefficients of which are gauge invariant.
The logarithmic divergences occur in these integrals, so the resummation of
these divergences does not influence the proof of gauge invariance.

In the case of the non-factorizable diagrams for a single unstable particle,
the resonant propagator now occurs inside a loop as
$1/\bigl((p+Q)^2-m^2\bigr)$, with $Q$ the integration momentum.  Because of
the infrared nature of the integral the main contribution is at $Q\approx0$
and we recover the original $1/(p^2-m^2)$, plus logarithms $\log(p^2-m^2)$, up
to the $n$-th power for an $n$ loop integral.  These logarithms again render
the residue at the pole $p^2-m^2$ undefined, so once more we decompose the
amplitude in a sum of fundamental loop integrals times coefficient functions.
Now, however, the linear divergence is also contained in the fundamental
integrals.  A gauge invariant separation of the resonant terms is therefore
accomplished by classifying the fundamental loop integrals into divergent and
non-divergent ones for $p^2\to m^2$.

The poles in the resonant fundamental loop integrals should of course resum to
the same $1/(p^2-M^2)$ as the propagator in the factorizable diagrams, with
the arguments of the logarithms also shifting into the complex plane.  Because
of the logarithmic terms, we cannot perform the expansion
(\ref{eq:mexpansion}) after integrating over the loops which comprise unstable
particles.  Just like in the case of the logarithmic on-shell divergences
encountered in section~\ref{subsec:charged}, the propagator can be resummed
{\em before\/} this integration is done (in the fundamental integrals).
Again, this expansion around the pole depends very much on the fact that this
is the applicable kinematical regime; thus using the on-shell width is only
valid when the integration momentum $Q$ is not too large.  As the singularity
is infrared in nature, this holds for the first order (one loop); for higher
orders more sophisticated techniques to isolate the poles will have to be
used.

In case of more than one unstable particle one again uses the same scheme for
each propagator separately.  The only difference is that now the
non-factorizable terms exhibit a much wider variety of singularities.  The
degree of divergence of the on-shell singularity obviously is equal to the
number of unstable particle propagators spanned, but the form is no longer the
simple product of propagators that occurs in the factorizable terms.  The
one loop integrals are enumerated below.


\subsection{One loop integrals}

\begin{figure}[htb]
\setscale{.8}
\unitlength .8bp
\begin{center}
\begin{picture}(150,120)(0,10)
\ArrowLine( 25, 25)( 50, 50)
\ArrowLine( 50, 50)( 50,100)
\ArrowLine( 50,100)( 25,125)
{\def\axowidth{1.5 }
\Photon(50, 50)(100, 50){3}{4}
}
\Photon(50,100)(100,100){3}{4}
\ArrowLine(125,125)(100,100)
\ArrowLine(100,100)(100, 50)
\ArrowLine(100, 50)(125, 25)
\Vertex( 50, 50){2}
\Vertex( 50,100){2}
\Vertex(100, 50){2}
\Vertex(100,100){2}
\put( 25, 25){\makebox(0,0)[br]{$e^-$}}
\put( 25,125){\makebox(0,0)[tr]{$e^+$}}
\put( 75, 45){\makebox(0,0)[t]{$Z$}}
\put( 75,105){\makebox(0,0)[b]{$\gamma$}}
\put(127, 25){\makebox(0,0)[bl]{$\mu^-$}}
\put(127,125){\makebox(0,0)[tl]{$\mu^+$}}
\put(125, 75){\makebox(0,0)[l]{$\Biggr)Z$}}
\put( 75,  0){\makebox(0,0)[b]{I}}
\end{picture}
\hfill
\begin{picture}(150,120)(0,10)
\Photon(50,100)( 25,125){3}{2.5}
\Photon(50, 50)(100, 50){3}{4}
{\def\axowidth{1.5 }
\Photon(50,100)(100,100){3}{4}
\Photon(25, 25)( 50, 50){-3}{2.5}
\Photon(50, 50)(50,100){3}{4}
}
\ArrowLine(125,125)(100,100)
\ArrowLine(100,100)(100, 50)
\ArrowLine(100, 50)(125, 25)
\Vertex( 50, 50){2}
\Vertex( 50,100){2}
\Vertex(100, 50){2}
\Vertex(100,100){2}
\put( 25, 25){\makebox(0,0)[br]{$W^{+}$}}
\put( 25,125){\makebox(0,0)[tr]{$\gamma^*,Z^*$}}
\put( 75, 45){\makebox(0,0)[t]{$\gamma$}}
\put( 75,105){\makebox(0,0)[b]{$W$}}
\put( 45, 75){\makebox(0,0)[r]{$W$}}
\put(125, 25){\makebox(0,0)[bl]{$\mu^-$}}
\put(125,125){\makebox(0,0)[tl]{$\bar{\nu}_\mu$}}
\put(125, 75){\makebox(0,0)[l]{$\Biggr)W^-$}}
\put( 75,  0){\makebox(0,0)[b]{II}}
\end{picture}
\hfill
\begin{picture}(150,120)(0,10)
\Photon(25,125)( 50,100){3}{2.5}
\ArrowLine(125,125)(100,100)
{\def\axowidth{1.5 }
\ArrowLine(50,100)(50, 50)
\ArrowLine(50, 50)( 25, 25)
\ArrowLine(100,100)( 50,100)
\Photon(125,25)(100, 50){-3}{2.5}
\Photon(100,50)(100,100){3}{4}
}
\Photon(50, 50)(100, 50){3}{4}
\Vertex( 50, 50){2}
\Vertex( 50,100){2}
\Vertex(100, 50){2}
\Vertex(100,100){2}
\put( 25,125){\makebox(0,0)[tr]{$\gamma^*,Z^*$}}
\put( 25, 25){\makebox(0,0)[br]{$t$}}
\put( 75,105){\makebox(0,0)[b]{$\bar{t}$}}
\put( 75, 45){\makebox(0,0)[t]{$\gamma$}}
\put( 45, 75){\makebox(0,0)[r]{$t$}}
\put(125,125){\makebox(0,0)[tl]{$\bar{b}$}}
\put(125, 25){\makebox(0,0)[bl]{$W^-$}}
\put(125, 75){\makebox(0,0)[l]{$\Biggr)\bar{t}$}}
\put( 75,  0){\makebox(0,0)[b]{III}}
\end{picture}
\end{center}
\caption{Examples of the three types of linearly divergent four point
functions.  The parentheses denote that the decay products are close to
resonance.}
\label{fig:Ds}
\end{figure}
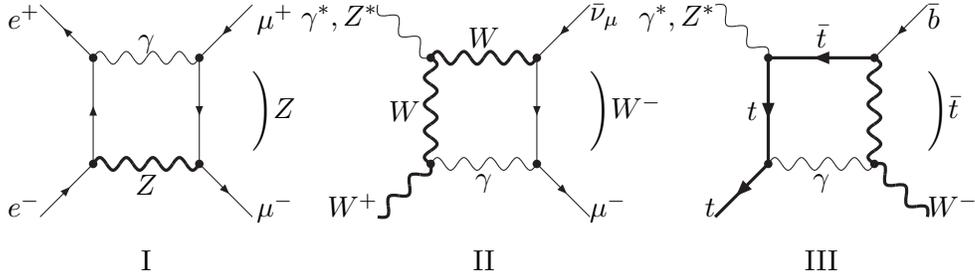

The first scalar function that can contain linear divergences is the four
point function.  Here we find three distinct singularity structures: one, two
and three unstable particles in the loop.\footnote{One needs one photon to
obtain the singularity.  There is one more divergent integral with no
unstable particle propagator but regulated by a small photon mass $\lambda$,
which is unphysical: it diverges as $1/\lambda$ (without logarithms).}
Examples of these are shown in Fig.~\ref{fig:Ds}.

The first integral has been given in Ref.\ \cite{Beenakker&DennerIR}, also for
a complex mass.  The structure for the assignment in Fig.~\ref{fig:Ds}I is
given by ($\lambda$ is the regulatory photon mass and $-4\delta_t =
(t-m_e^2-m_\mu^2)^2 - 4m_e^2m_\mu^2$ the K\"all\'en function)
\begin{equation}
D_0^{\rm I} = \frac{1}{M_Z^2-s}
    \frac{1}{2\sqrt{-\delta_t}}
    \Biggl\{ 2\log\biggl(
        \frac{t-(m_\mu-m_e)^2 - 2\sqrt{-\delta_t}}
             {t-(m_\mu-m_e)^2 + 2\sqrt{-\delta_t}} \biggr)
        \log\biggl(\frac{\sqrt{M_Z^2}\lambda}{M_Z^2-s}\biggr)
    + \makebox{finite terms} \Biggr\}
\end{equation}
The introduction of the finite width does not pose any problems.

The second four point function is more difficult to get correctly.  The
structure is
\begin{eqnarray}
    D_0^{\rm II} & = &
         \frac{1}{\sqrt{\delta^{s_1 s_2 s_3 s_4}_{s_1 s_2 s_3 s_4}}}
         \Biggl\{ \sum\Li\left(f_i\Bigl(\frac{(s_1 s_3)}{(s_1 s_4)}
            \Bigr)\right)
         \Biggr\}
    \;,
\label{eq:nastyD0}
\end{eqnarray}
with $(s_1 s_3) = (m_W^2 - p_+^2)/2$, $(s_1 s_4) = (m_W^2 - p_-^2)/2$ and the
determinant given by
\begin{eqnarray}
    \delta^{s_1 s_2 s_3 s_4}_{s_1 s_2 s_3 s_4} & = & \left|
        \begin{array}{ccc} (s_1 s_1) & \cdots & (s_1 s_4) \\
                              \vdots &        & \vdots \\
                           (s_4 s_1) & \cdots & (s_4 s_4) \end{array} \right|
\\
    (s_i s_j) & = & (m_i^2 + m_j^2 - p_{ij}^2)/2
\\
    (s_i p_{jk}) & = & (m_j^2 - m_k^2 - p_{ij}^2 + p_{ik}^2)/2\;,
\end{eqnarray}
the momentum $p_{ij}$ being defined as the difference of the momenta flowing
through propagators with masses $m_i$ and $m_j$.  The function $f_i$ are
complicated functions of their arguments and the other on-shell invariants.

The third type does not occur in $W$ pair production and we have not yet
investigated it.

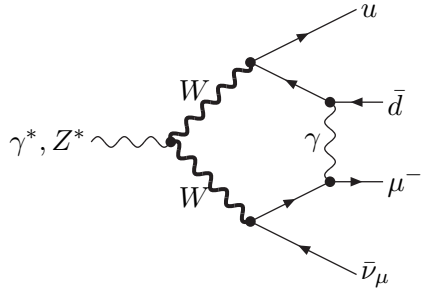
\begin{figure}[htb]
\begin{center}
\begin{picture}(120,110)(0,10)
\Photon(0,60)(30,60){2}{3}
{\def\axowidth{1.5 }
\Photon(30,60)(60,90){2}{5}
\Photon(30,60)(60,30){2}{5}
}
\ArrowLine(110,75)(90,75)
\ArrowLine(90,75)(60,90)
\ArrowLine(60,90)(100,110)
\ArrowLine(100, 10)(60,30)
\ArrowLine(60,30)(90,45)
\ArrowLine(90,45)(110,45)
\Photon(90,45)(90,75){2}{3}
\Vertex(30,60){2}
\Vertex(60,90){2}
\Vertex(60,30){2}
\Vertex(90,45){2}
\Vertex(90,75){2}
\put(-2,60){\makebox(0,0)[r]{$\gamma^*,Z^*$}}
\put(45,76){\makebox(0,0)[br]{$W$}}
\put(45,43){\makebox(0,0)[tr]{$W$}}
\put(112, 75){\makebox(0,0)[l]{$\bar{d}$}}
\put(102,110){\makebox(0,0)[l]{$u$}}
\put(112, 45){\makebox(0,0)[l]{$\mu^-$}}
\put(102, 10){\makebox(0,0)[l]{$\bar{\nu}_\mu$}}
\put(87, 60){\makebox(0,0)[r]{$\gamma$}}
\end{picture}
\end{center}
\caption{Example of a quadratically divergent five point function}
\label{fig:Es}
\end{figure}

The five point function can maximally contain a quadratic on-shell divergence.
If it is less singular, we just decompose it into five four point functions
\cite{VanNeerven&Vermaseren5point,GJthesis} and apply the results given above.
An example of a quadratically divergent five point function in $W$ pair
production is shown in Fig.~\ref{fig:Es}.\footnote{Again, in the physically
uninteresting case that the all particles are taken on-shell and the
divergence is regulated by a small photon mass the $E_0$ diverges as
$1/\lambda^2$.}  In a quadratically divergent five point function the extra
singularity occurs in the determinants multiplying the four point functions in
the general expression
\begin{equation}
    E_0 = \frac{-1}{2\delta^{s_1 s_2 s_3 s_4 s_5}_{s_1 s_2 s_3 s_4 s_5}}\Bigg\{
        \sum_{i=1}^5
\delta{}^{s_i}_{p_1}{}^{s_{i+1}}_{p_2}{}^{s_{i+2}}_{p_3}{}^{s_{i+3}}_{p_4} \;
        D_0(i,i+1,i+2,i+3) \Biggr\},
\label{eq:E0}
\end{equation}
with the determinants
\begin{eqnarray}
    \delta^{s_1 s_2 s_3 s_4 s_5}_{s_1 s_2 s_3 s_4 s_5} & = & \left|
        \begin{array}{ccc} (s_1 s_1) & \cdots & (s_1 s_5) \\
                              \vdots &        & \vdots \\
                           (s_5 s_1) & \cdots & (s_5 s_5) \end{array} \right|
\\
    \delta{}^{s_i}_{p_1}{}^{s_{i+1}}_{p_2}{}^{s_{i+2}}_{p_3}{}^{s_{i+3}}_{p_4}
        & = & \left|
        \begin{array}{ccc} (s_i p_1) & \cdots & (s_i p_4) \\
                              \vdots &        & \vdots \\
                       (s_{i+3} p_1) & \cdots & (s_{i+3} p_4) \end{array}
              \right|
    \;.
\end{eqnarray}
$D_0(i,j,k,l)$ is the four point function with propagators with masses $m_i,
m_j,m_k$ and $m_l$.  We have to use complex masses in the overall factor,
which is quadratically divergent.  This is reduced to a linear divergence by
the coefficient determinant $\delta \strut^{s_i}_{p_1} \strut^{s_{i+1}}_{p_2}
\strut^{s_{i+2}}_{p_3} \strut^{s_{i+3}}_{p_4}$ in the four terms with a
linearly divergent four point function.  In the fifth term, where the photon
is not included in the $D_0$, both the determinant and the four point function
are regular, thus retaining the quadratic divergence. Note that, for
consistency, we need a complex mass even in this last four point function,
even if its main contribution is not at $Q=0$.\footnote{Unless a corresponding
modification is made in the soft bremsstrahlung integrals.}  This integral
$\int d^4Q/\bigl[ (p_{12}Q)(p_{13}Q)(p_{14}Q)(p_{15}Q) \bigr]$ is regular,
even though power counting would indicate a non-standard logarithmic
divergence at $Q=0$.

All other higher point function are treated similarly.  As we have had no
immediate use for the cubically divergent six point function we have not
analyzed it in detail.  The quadratically divergent $F_0$ is trivially
decomposed in six five point functions.


\subsection{Soft bremsstrahlung}

As was noted above, the soft bremsstrahlung integrals ($E_\gamma < \omega \ll
\Gamma$) have the same resonance structure as the virtual ones.   The
treatment given above can thus be taken over completely.  It has been argued
that the resonant part of the non-factorizable virtual diagrams (times lowest
order) is equal and opposite to that of the corresponding soft bremsstrahlung
graphs \cite{KhozeRad}, so that the sum is non-resonant and can be neglected.
This effect is also known from the $Z$ pole \cite{BeenakkerThesis}.  However,
as in the case of the logarithmic divergences, this cancellation only occurs
when one integrates out the soft photons up to an energy $\omega \gg \Gamma$.
Again, one cannot integrate when these small energy scales are relevant; of
course if they are not important the narrow width approximation would have
been sufficient.



\section{Gauge invariance}
\label{sec:gauge}

In the previous sections we have resummed the factorizable resonant and
non-factorizable resonant graphs in a complicated fashion.  This gave rise to
an amplitude of the form (for a single unstable particle)
\begin{equation}
    {\cal A} = \frac{w}{p^2-M^2} + n(p^2)
\;,
\end{equation}
where $M^2$, $w$ and $n(p^2)$ are computed from the analytic properties of the
unresummed amplitude.  Now we show, essentially following M.~Veltman
\cite{VeltmanUnstable}, that this way of resummation does in fact give a gauge
invariant result.  Gauge invariance here denotes invariance with respect to
any gauge group --- the arguments are very general and only depend on the
assumption that a fixed order amplitude is gauge invariant for all $p^2$,
which is used in the limit $p^2\to m^2$ (where the fixed-order calculation is
not applicable for physics results).

The first point to note is that the pole position $M^2$ is a property of the
$S$ matrix and a physically measurable quantity; it should therefore be gauge
invariant.  In this scheme it is given by the solution of the equation
$M^2 - m^2 - \Pi(M^2) = 0$.

The residue at the pole $w$ is given as a sum of the resonant contributions
from the factorizable and non-factorizable diagrams.  However, the origins of
the poles are quite different: in the non-factorizable diagrams it is provided
by linearly (or higher) divergent fundamental loop integrals, in the
factorizable diagrams by explicit propagators in the coefficients of these
integrals.  As the fundamental loop integrals are linearly independent, the
resonant part of the non-factorizable diagrams, being composed of different
fundamental integrals, must be separately gauge invariant to each order.

For the factorizable part it is easiest to look at the unresummed amplitude
truncated at a fixed order $n$, which is gauge invariant.  In the (unphysical)
limit $p^2\to m^2$ this diverges as $1/(p^2-m^2)^{n+1}$; the coefficient of
this pole $W_{-n-1}$ is therefore gauge invariant.  Subtracting this pole and
repeating the procedure for the remainder we find that all $W_{-i}$ are gauge
invariant, in particular $W_{-1}$, which gives the factorizable contribution
to the residue in this order.  This part therefore is also order-by-order
gauge invariant.  The occurrence of divergent logarithms does not spoil this
argument due to the fact that they only occur inside the fundamental loop
integrals.

Finally, because the resonant parts of the unresummed amplitude are all gauge
invariant, so is the sum of all the non-resonant contributions in the resummed
expression.


\section{Summary}

We have given a gauge invariant prescription to treat unstable particles in
loop calculations, with explicit results for the one loop case.  The scheme is
an extension of the pole expansion method of M.~Veltman
\cite{VeltmanUnstable,Stuart1,HVeltmanUnstable} to include several charged
resonances and loops.   An advantage is also that only a minimal subset of
diagrams is needed.  It is shown that all necessary quantities can be obtained
from a normal perturbative off-shell expansion.  The problems which arise in
the case of charged resonances with infrared divergences have all been
addressed, as well as the corresponding treatment of the soft bremsstrahlung
integrals ($E_\gamma < \omega \ll \mbox{width}$).

\smallskip

\noindent For a tree level calculation in this scheme one needs
\begin{itemize}
\item the on-shell width $\Pi^{(1)}(M^2)$ to one loop and
\item the resonant tree level amplitude $W^{(0)}(m^2)$.
\end{itemize}
The $\order(\alpha)$ corrections are given by:
\begin{itemize}
\item improving the on-shell width $\Pi^{(2)}(M^2)$ to two loops,
\item the one loop factorizable diagrams $W^{(1)}(m^2)$ with the substitution
$m^2\to M^2$ in the divergent scalar three point functions,
\item the resonant part of the one loop non-factorizable graphs,
\item the derivative terms $W^{(0)}(m^2)\Pi(m^2)$ and
$W^{(0)}{}'(m^2)\Pi(m^2)$,
\item the tree level $p^2$ dependent resonant terms $W^{(0)}(p^2)$ and
\item the tree level non-resonant terms.
\end{itemize}
The soft bremsstrahlung integrals are treated analogously.

\smallskip

The resonance structure of the non-factorizable graphs (those with unstable
particles in the loops) is quite complicated in the case of multiple
resonances (like $W$ pair production): they are not just products of poles.
These give a contribution which is not suppressed by additional powers of the
width.   The corresponding soft bremsstrahlung integrals only cancel this
contribution if one integrates out the photons up to an energy much larger
than the width.

The expansion around the poles underlying our scheme leads to difficulties
near (especially below) the thresholds for the production (or decay) of the
unstable particles.  Connecting the regions above and below threshold will
require other techniques.

\paragraph{Acknowledgements.}
We would like to thank Fred Jegerlehner, Wolfgang Hollik, Thomas Sack, Andreas
Salath\'e and Jos Vermaseren for many helpful discussions.  One of us (GJvO)
would like to thank FermiLab for the hospitality while debugging the complex
functions.


\bibliographystyle{nucphys}
\bibliography{fenomeen}


\appendix
\section{Proof of Eqs (\protect\ref{eq:w=W}) and (\protect\ref{eq:n=N})}
\label{ap:proofs}

In order to prove Eq.\ (\ref{eq:w=W}) we write (writing $s=p^2$)
\begin{equation}
    W_{-1} =
      \sum_{n=0}^\infty\frac{1}{n!}\,\frac{d^n}{ds^n}
      \left[ W(s)\Pi^{n}(s) \right]_{s=m^2}.
\end{equation}
We reexpress each term of the sum by expanding around $s=M^2$ and using
$m^2-M^2=-\Pi(M^2)$
\begin{eqnarray}
    & = & \sum_{n=0}^\infty \frac{1}{n!} \sum_{k=0}^\infty \frac{1}{k!}\;
       \frac{d^{n+k}}{ds^{n+k}}
      \left[ W(s)\Pi^{n}(s) \right]_{s=M^2} \;\bigl(-\Pi(M^2)\bigr)^k
\nonumber\\
   & = & \sum_{n,k=0}^\infty \frac{1}{(n+k)!}\;
      {n+k \choose n} \frac{d^{n+k}}{ds^{n+k}}
      \left[ W(s)\Pi^{n}(s) \bigl(-\Pi(M^2)\bigr)^k \right]_{s=M^2}
\nonumber\\
    & = & \sum_{n'=0}^\infty \frac{1}{n'!}\;\frac{d^{n'}}{ds^{n'}}
      \left[ W(s)\bigl(\Pi(s)-\Pi(M^2)\bigr)^{n'} \right]_{s=M^2}
\end{eqnarray}
with $n' = n+k$.  Now we can expand $\Pi(s)$ as well as $W(s)$ around $s=M^2$.
Only the terms without any $(s-M^2)$-factors will survive; therefore we get
\begin{equation}
     = \sum_{n'=0}^\infty W(M^2)(\Pi'(M^2))^{n'} =
        \frac{W(M^2)}{1-\Pi'(M^2)} = w.
\end{equation}

The proof of the second relation (\ref{eq:n=N}) follows along similar lines.
We start by writing $N(s)$ as
\begin{eqnarray}
    N(s) & = & \sum_{\ell=1}^\infty (s-m^2)^{\ell-1}
        \sum_{n=0}^\infty \frac{1}{(n+\ell)!} \frac{d^{n+\ell}}{ds^{n+\ell}}
        \left[ W(s)\Pi^n(s) \right]_{s=m^2}
\nonumber\\
         & = & \sum_{\ell=1}^\infty (s-m^2)^{\ell-1}
        \left\{ \sum_{n'=0}^\infty - \sum_{n'=0}^{\ell-1} \right\}
        \frac{1}{n'!} \frac{d^{n'}}{ds^{n'}}
        \left[ \frac{W(s)}{\Pi^\ell(s)}\;\Pi^n(s) \right]_{s=m^2}
\end{eqnarray}
with $n' = n+\ell$.  Using the previous proof the first term in braces gives
\begin{equation}
    \sum_{\ell=1}^\infty (s-m^2)^{\ell-1} \frac{W(M^2)}{\Pi^\ell(M^2)}
        \frac{1}{1-\Pi'(M^2)}
    = - \frac{W(M^2)}{s-M^2} \frac{1}{1-\Pi'(M^2)}\;.
\end{equation}
The second term is a simple Taylor series after changing the summation index
to $\ell' = \ell - n'$:
\begin{eqnarray}
    \lefteqn{ - \sum_{\ell'=1}^\infty \sum_{n'=0}^\infty (s-m^2)^{\ell'+n'-1}
    \frac{1}{n'!} \frac{d^{n'}}{ds^{n'}}
        \left[ \frac{W(s)}{\Pi^{\ell'}(s)} \right]_{s=m^2} }
\nonumber\\
    & = & - \sum_{\ell'=1}^\infty \sum_{n'=0}^\infty (s-m^2)^{\ell'-1}
        \frac{W(s)}{\Pi^{\ell'}(s)} = \frac{W(s)}{s-m^2-\Pi(s)}\;.
\end{eqnarray}
The sum is thus exactly $n(s)$, which was to be shown.


\section{Frame dependence}
\label{ap:frame}

We will show that, up to terms of $\order(\alpha^2)$, Eq.\ \ref{eq:Mexpansion}
gives a result which is independent of the definition of the other kinematical
arguments of $W$.  In particular, when these are angles, the result will only
depend on the frame chosen through terms proportional to $\order(\alpha^2)$.

Let us compute first in one frame, in which $W$ is a function of $p^2$ and
another set of variables $x_i$.  The contribution of the factorizable diagrams
than reads, up to $\order(\alpha)$,
\begin{equation}
    {\cal A}^{(0+1)}_{\rm fact} = \frac{W(p^2,x_i)-W(m^2,x_i)}{p^2-m^2}
        + \frac{W(m^2,x_i) + \frac{d}{dp^2}\left.W(p^2,x_i)\Pi(p^2)
            \right|_{p^2=m^2}}{p^2-M^2}
\;.
\end{equation}
In another frame we have $x_i = x_i(y_i,p^2)$, where the $y_i$ denote the
kinematical variables in that frame, and the dependence of the Lorentz
transformation on $p^2$ is shown explicitly.  Suppressing the $y_i$ the
amplitude now is
\begin{equation}
    {\cal A'}^{(0+1)}_{\rm fact} = \frac{W(p^2,x_i(p^2))-W(m^2,x_i(m^2))}
        {p^2-m^2} + \frac{W(m^2,x_i(m^2)) + \frac{d}{dp^2}\left.
        W(p^2,x_i(p^2))\Pi(p^2)\right|_{p^2=m^2}}{p^2-M^2}
\;.
\end{equation}
Substituting back $x_i = x_i(p^2)$ the difference is given by
\begin{eqnarray}
    {\cal (A'-A)}^{(0+1)}_{\rm fact} & = &
        \frac{W(m^2,x_i(p^2))-W(m^2,x_i(m^2))}{p^2-m^2} + \frac{W(m^2,x_i(m^2))
        - W(m^2,x_i(p^2))}{p^2-M^2}
\nonumber\\&&\mbox{}
     + \frac{\frac{\partial W}{\partial p^2}(m^2,x_i(m^2))
        - \frac{\partial W}{\partial p^2}(m^2,x_i(p^2)) + \sum_i
        \frac{\partial W}{\partial x_i} \frac{\partial x_i}{\partial p^2}
        (m^2,x_i(m^2))}{p^2-M^2}
\;.
\end{eqnarray}
Expanding this around $p^2=m^2$ and using $M^2 = m^2 + \Pi(m^2) +
\order(\alpha^2)$ this leaves only terms of order $\alpha^2$ and
$(p^2-m^2)\alpha$.


\section{Gauge invariance of the coefficients of scalar functions}
\label{ap:scalargauge}

The fact that the coefficients of the scalar functions must be gauge invariant
can be seen for the electroweak section of the standard model when we consider
it with masses for all particles (except the photon and gluon).  This does not
break gauge symmetry.  One can then consider the scalar functions in turn, and
consider whether they can be linearly dependent.  Note that this implies that
the coefficients of scalar functions involving gauge-dependent parameters (for
instance ghost masses) must vanish.

\begin{itemize}
\item The one point function
\begin{equation}
A_0(m_i^2) = -m_i^2 \log(m_i^2) + \ldots
\end{equation}
is obviously linearly independent of any other one point functions.
\item The two point functions
\begin{equation}
B_0(m_i^2,m_j^2,p_k^2) = \frac{2\sqrt{-\delta(m_i^2,m_j^2,p_k^2)}}
    {2p_k^2} \log\Bigl( \frac{m_i^2-m_j^2+p_k^2+2\sqrt{-\delta}}
    {m_i^2-m_j^2+p_k^2-\sqrt{\lambda}} \Bigr) + \ldots
\end{equation}
is unique for $i \leq j$ and non-zero masses as the factor $\sqrt{-\delta}$
does not occur in the one point functions or coefficients of other two point
functions (except with permuted arguments, in which case the logarithms are
different).  The scalar two point functions with $p_k^2=0,\; m_i^2=m_j^2$
or $m_i^2=p_k^2, \; m_j^2=0$ reduce to one point functions; the remaining
possibility has a unique logarithm.
\item The three point function
\begin{equation}
C_0(m_i^2,m_j^2,m_k^2,p_l^2,p_m^2,p_n^2) =
\frac{1}{2\sqrt{-\delta(p_l^2,p_m^2,p_n^2)}} \sum_{\ell=1}^6 \Li(c_\ell)
\end{equation}
must again be brought into a unique form $i \leq j \leq k$ (if equal, order
$l,m,n$).  In case the kinematical determinant $\delta(p_l^2,p_m^2,p_n^2)$ is
not equal to zero, it contains dilogs (or double logs), which are independent
of any one or two point functions.  The overall factor, again with a root,
uniquely identifies the momenta; but different three point functions with the
same external momenta but different masses will have different arguments for
the dilogarithms.  When the kinematical determinant is zero this no longer is
the case, and three point functions do in fact reduce to two point functions
\cite{FleischerJegerlehnerHiggs1}.  Except for threshold (which we do not
consider) these occur in counterterms (for instance $\gamma WW$ at $q^2=0$).
\item The four point function depends on 6 masses and 10 momenta squared.
It has the form
\begin{equation}
D_0 = \frac{1}{\sqrt{\delta^{s_1 s_2 s_3 s_4}_{s_1 s_2 s_3 s_4}}}
    \sum_{\ell=1}^{16} \Li(c_\ell)
\end{equation}
with the $s_i$ internal momenta defined before.  The overall factor has the
same symmetries as the $D_0$ and depends on all the masses and momenta squared
in the box, so after bringing it to a standard order it uniquely identifies
the four point function when it is non-zero.  There are two cases where it is
zero: real singularities on the boundaries of phase space ($\delta^{p_1 p_2
p_3}_{p_1 p_2 p_3} = 0$; this again occurs on threshold and in counterterms)
and artificial ones within phase space.  The latter ones depend on the values
for $s$ and $t$, at least one of which can normally be chosen freely.  It can
therefore not lead to a dependency which is valid for all allowed $s$ and $t$.
\item
The five point function is a linearly dependent set of five four point
functions.  However, this decomposition crucially depends on the
dimensionality of space (it uses the Schouten identity).  As the gauge
invariance of the amplitude does not depend on this, the coefficient of the
$E_0$ must be gauge invariant on its own.  The same holds for six- and
higher point functions.
\end{itemize}
We have thus shown that all scalar functions with non-zero kinematical
determinants are linearly independent.  The others should be reduced to lower
point functions.

\end{document}